\documentclass[preprint]{vgtc}                          

\graphicspath{{figures/}{pictures/}{images/}{./}} 

\usepackage{times}                     

\usepackage{tabu}                      
\usepackage{booktabs}                  
\usepackage{lipsum}                    
\usepackage{mwe}                       
\usepackage{soul}
\usepackage{balance}

\usepackage{mathptmx}                  

\onlineid{0}

\vgtccategory{Research}

\vgtcinsertpkg

\preprinttext{To appear in Proceedings of the 2024 IEEE Visualization Conference ({VIS '24})}

\title{Science in a Blink: Supporting Ensemble Perception in Scalar Fields}

\author{Victor A. Mateevitsi\thanks{e-mail: vmateevitsi@anl.gov}\\
\parbox{1.5in}{\scriptsize \centering Argonne National Laboratory \\ University of Illinois Chicago}
\and Michael E. Papka\thanks{e-mail: papka@anl.gov}\\
\parbox{1.5in}{\scriptsize \centering Argonne National Laboratory \\ University of Illinois Chicago}
\and Khairi Reda\thanks{e-mail: redak@iu.edu.}\\\scriptsize Indiana University Indianapolis}

\teaser{
  \centering
  \includegraphics[width=1\linewidth]{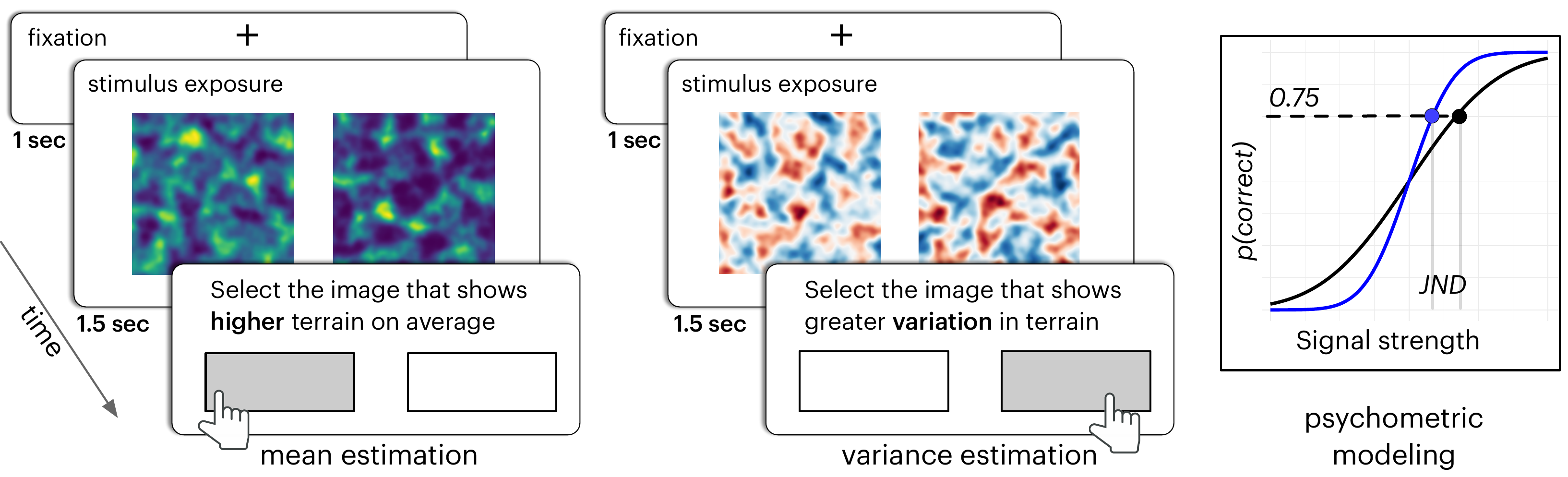}
  \caption{A typical trial sequence. Participants were asked to rapidly (after 1.5 seconds exposure) estimate one of two ensemble statistics, the \emph{mean and variance}, from scalar field pairs. They indicated their response by selecting the side exhibiting the higher statistic. We model responses as the likelihood of identifying the correct target subject to colormap and the signal difference between the two fields.}
  \label{fig:stimulus}
}

\abstract{Visualizations support rapid analysis of scientific datasets, allowing viewers to glean aggregate information (e.g., the mean) within split-seconds. While prior research has explored this ability in conventional charts, it is unclear if spatial visualizations used by computational scientists afford a similar \emph{ensemble perception} capacity. We investigate people's ability to estimate two summary statistics, mean and variance, from pseudocolor scalar fields. In a crowdsourced experiment, we find that participants can reliably characterize both statistics, although variance discrimination requires a much stronger signal. Multi-hue and diverging colormaps outperformed monochromatic, luminance ramps in aiding this extraction. Analysis of qualitative responses suggests that participants often estimate the distribution of hotspots and valleys as visual proxies for data statistics. These findings suggest that people's summary interpretation of spatial datasets is likely driven by the appearance of discrete color segments, rather than assessments of overall luminance. Implicit color segmentation in quantitative displays could thus prove more useful than previously assumed by facilitating quick, gist-level judgments about color-coded visualizations.
} 


\begin{document}


\firstsection{Introduction}

\maketitle

Scalar fields are a ubiquitous data type appearing in various disciplines~\cite{zeller2020visualizing}, including climate and earth science, physics, and materials science. In visualizing these fields, scientists typically employ a smooth color scale to map scalar values to continuous color gradations. Several guidelines have been proposed to enhance the perception of colormaps for scalar data, such as ensuring perceptual uniformity and smoothness~\cite{zhou2016survey}, inspired by empirical evaluations~\cite{rogowitz1999trajectories}. However, empirical studies have primarily focused on tasks that involve reading localized information from fields, such as retrieving the scalar value at a specific point on a map~\cite{ware1988color,reda2018graphical} or detecting small features~\cite{ware2018measuring}. Although informative, these tasks do not fully encompass the breadth of use cases. For instance, when comparing two fields representing alternative models or simulation runs, a scientist is less likely to rely on point estimates or individual features alone. Instead, comparisons often entail a holistic assessment, wherein the scientist might compare aggregate properties (e.g., variance) from across multiple visualizations. These summary properties, sometimes referred to as \emph{ensemble representations}~\cite{alvarez2011representing}, reflect a rapid pooling of features by the visual system, allowing one to render quick, `gist' assessments. This in turn allows a scientist to answer questions like which simulation exhibits a higher mean temperature or greater variance in fluid velocity. Ensemble perception could allow for such questions to be answered within seconds of looking at visualizations.

Researchers have explored how ensemble representations allow for efficient extraction of summary data from visualizations~\cite{szafir2016four}. These studies insofar involved conventional charts like scatterplots, typically evaluating the accuracy of perceiving the average point position~\cite{gleicher2013perception} or size~\cite{lau2018ensemble}. However, research has yet to characterize the ensemble-coding affordances of spatial displays utilized by the computational science community, like scalar fields. Such visualizations employ color as the primary encoding channel, giving rise to different visual features than the discrete marks found in charts. Whether and how well people might be able to extract aggregate data properties from scientific visualizations carries important implications for the analysis and communication of scientific data (e.g., weather predictions and fire hazard maps). Moreover, because the nature of features in these visualizations will differ depending on the colormap, it is vital to understand how well different color designs support summary inference. 

To address these questions, we conducted a crowdsourced experiment evaluating people's perceptions of two aggregate statistics, the mean and variance. Participants were asked to estimate either statistic from scalar fields with brief exposure times, thus stimulating rapid ensemble perception processes. 
We find that participants could estimate both statistics, with mean discrimination superior and requiring a smaller signal. Multi-hue, diverging, and rainbow colormaps outperformed simple greyscale ramps. These results support the hypothesis that ensemble perception in spatial visualizations involves estimating discrete color features (e.g., distribution of colors representing peaks and valleys in the data). The results offer insights for designing colormaps to facilitate rapid visual communication of scientific data.

\section{Related Work}

\subsection{Ensemble Perception}
Substantial evidence indicates that the visual system can quickly perceive global properties of sets of items~\cite{whitney2018ensemble,alvarez2011representing}. Ariely shows that people can quickly estimate the average  size for a set of circles with a glance, even when they are unable to recall individual objects~\cite{ariely2001seeing}. Subsequent studies have replicated and expanded these results, showing that the visual system quickly encodes not just sizes, but also the mean orientation~\cite{haberman2015individual}, position~\cite{gleicher2013perception}, and color~\cite{webster2014perceiving,virtanen2020color} for collections. Remarkably, this ability extends to higher-level features like the average facial emotion by a crowd~\cite{haberman2007rapid}. Beyond encoding the central tendency, the visual system also appears to estimate higher-order statistics like variance or skewness~\cite{morgan2008dipper,lau2018ensemble}.

Although the precise mechanism behind ensemble perception is still debated~\cite{myczek2008better,im2013effects}, research suggests the involvement of distinct perceptual processes than those used for focused attention~\cite{baek2020ensemble}, allowing for global, summary information to be extracted in less than a second~\cite{sweeny2015ensemble}. These mechanisms can be exploited in visualizations~\cite{cui2021synergy, szafir2016four}: By using `channels' that support ensemble coding, we enable visualization viewers to quickly extract summary statistics about the underlying data. This includes the mean value in a timeseries~\cite{albers2014task} or the variance in a heatmap~\cite{cui2021synergy}. Prior work has primarily studied these averaging processes in InfoVis-style representations. It is unclear whether the same results extend to visualizations of `scientific' and spatial datasets. Whereas ensemble processes are thought to operate over collections of discrete objects or marks~\cite {franconeri2009number}, data types like scalar fields give rise to a contiguous, smooth representation with no clear notion of `items' or collections.  

\subsection{Colormaps for Scalar Data}

Guidelines suggest that quantitative colormaps should comprise a perceptually uniform ramp, maintaining even perceptual distances~\cite{rogowitz1999trajectories,rogowitz1998data,zhou2016survey}. Color ramps should also exhibit smooth gradations with no abrupt changes or boundaries in color~\cite{borland2007rainbow}, which could be misconstrued as false features. Although widely adopted in tools~\cite{salvi2024colormaker}, this guidance assumes that viewers will mentally translate differences in color to quantitative differences when interpreting the display. Recent studies, however, challenge this assumption, showing that colorful, non-uniform maps like rainbow can be advantageous when interpreting scalar fields~\cite{reda2020rainbows,reda2022rainbow,reda2021color}.
Nevertheless, it is unclear whether this advantage for rainbows extends to tasks requiring fast extraction of aggregate statistics from fields. Work by Warden et al. indicates that large hue variation can disrupt ensemble perception in visualizations~\cite{warden2022visualizing}. Instead, they suggest sequential ramps to support the perception of trends in strip plots. Luminance is indeed a feature that the visual system appears capable of averaging~\cite{bauer2009does}. Accordingly, a viewer could estimate the mean value in a scalar field or heatmap by gauging the average `brightness' of the display. Such a strategy might favor a sequential (or monochromatic) colormap with monotonically increasing luminance. Alternatively, the viewer might resort to assessing the number or size of peaks as a proxy for the mean -- the latter would be aided by a diverging or rainbow colormap. It is unclear which of these two strategies people follow. One study shows that gradient perception benefits from rainbows, although without controlling for visualization exposure time~\cite{reda2019evaluating}. However, gradients reflect local variation, as opposed to a global ensemble statistic. In this paper, we focus on two global features the visual system is known to extract and represent in extremely short duration. We investigate the effectiveness of different colormap designs for supporting this type of ensemble perception. Simultaneously, we identify which of the above feature categories, luminance vs. discrete color patches, observers draw upon to estimate the mean and variance in fields.

\begin{figure*}
    \centering
    \includegraphics[width=0.8\linewidth]{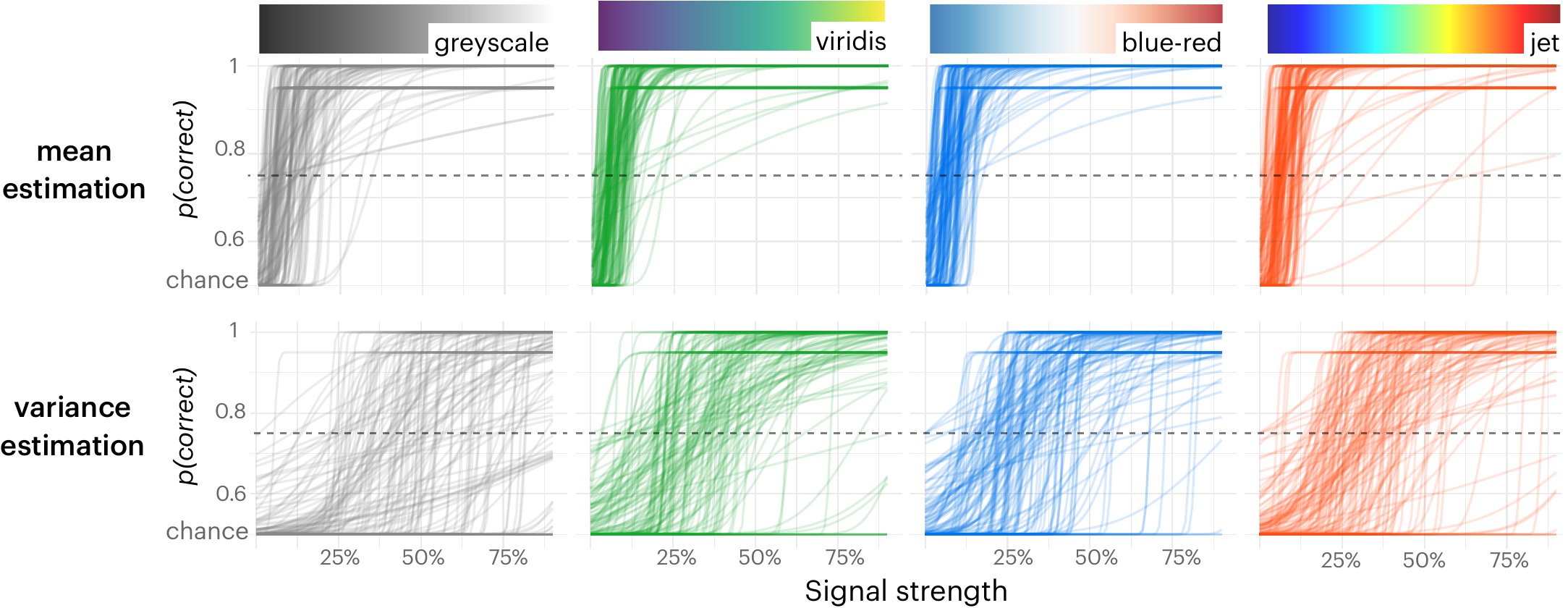}
    \caption{Psychometric functions were fitted to model participants' responses. The horizontal intercept (dashed line) indicates 75\% accuracy. Intersecting this line with the function gives the just noticeable difference (JND) threshold for each \emph{statistic} $\times$ \emph{participant} $\times$ \emph{colormap} combination.}
    \label{fig:psyc_functions}
    \vspace{-4mm}
\end{figure*}

\section{Research Questions \& Hypotheses}
\label{sec:hypotheses}

We address three research questions: 

\vspace{1mm}\noindent\textbf{RQ1:} To what extent can observers make rapid, summary judgments about the mean and variance in color-coded fields? How discriminable are these two statistical properties given a brief exposure time? We quantify the required signal strength for reliable judgments using psychometric modeling techniques~\cite{wichmann2001psychometric}.

\vspace{1mm}\noindent\textbf{RQ2:} Does the choice of colormap impact ensemble estimates? If so, which colormaps are most effective for conveying the mean vs. variance? We compare discrimination thresholds under various colormaps to measure their suitability for conveying the two statistics.

\vspace{1mm}\noindent\textbf{RQ3:} When making rapid summary judgments about fields, what visual features do observers rely upon? We test two hypotheses and analyze self-reported strategies to understand visual `proxies' people use to estimate statistical properties.


\vspace{3mm}\noindent\textbf{Hypotheses:} We posit two complementary (though non-mutually exclusive) hypotheses:

\vspace{1mm}\noindent\textbf{H1:} Participants will assess the overall luminance of the display to estimate the mean or variance of scalar fields. This assumption follows work showing that average `brightness' is an ensemble feature that can be rapidly extracted by the visual system~\cite{rajendran2021ensemble}. This hypothesis favors sequential colormaps (e.g., a plain greyscale or a multi-hue ramp like \emph{viridis}) over more colorful designs (e.g., diverging and rainbow) where luminance cannot predict data values.

\vspace{1mm}\noindent\textbf{H2:} Participants will infer the mean/variance by assessing the size and distribution of color segments (e.g., peaks and valleys). These features can serve as a visual indicator of how elevated or varied the scalar values are. This hypothesis favors colorful designs (e.g., \emph{jet}, \emph{red-blue}, and to a lesser extent \emph{viridis}) as the latter will introduce `useful' artifacts in the form of color bands~\cite{quinan2019examining}. Although seemingly problematic~\cite{borland2007rainbow}, we hypothesize that these features could help in estimating the statistical properties of fields at a glance.

\section{Experiment}

We conduct a crowdsourced experiment utilizing a two-alternative forced choice (2AFC) task~\cite{tyler2000signal}. Participants compared two scalar fields side-by-side and indicated which of the two exhibited a higher overall mean or variance. To restrict the experiment to measuring rapid ensemble processes, the scalar fields were only shown for a brief period of 1.5 seconds (similar to experiments on ensemble perception~\cite{lau2018ensemble}). After this short exposure time, the display was cleared and participants were prompted to make their selection. 

\subsection{Stimulus Generation}

We procedurally synthesize scalar fields ($250 \times 250$ pixels each) using a Perlin process~\cite{perlin1985image}. Each field is synthesized as follows: 

\begin{equation}
 I(x,y) = I(\vec{u}) = \sum_{i=1}^{m} \omega_i {F(2^{i-1} . \vec{u})}^k
\end{equation}

Where $\vec{u}=(x, y)$ is the image coordinates, and $F(x, y)$ represents the output of a 2D Perlin function. The summation blends multiple ($m=5$) octaves of noise with weights $\omega_i$. The exponent $k$ controls the distribution of scalar values. To obtain stimuli covering a range of mean and variance levels, we generated a sample of 10,000 fields while varying the exponent  $k$ between $[0, 3]$. We then compute the ensemble mean and variance by averaging $I(x,y)$ and computing the standard deviation. We selected a subset of the generated field to be used as stimuli such that there is no covariance between the two statistics. This allows us to systematically vary one of the two ensemble statistics independently from the other, ensuring that judgments are not confounded. Fields were presented to participants as height maps with color depicting terrain elevation.

\subsection{Participants, Experiment Design, and Procedures}

\noindent\textbf{Participants: }We recruited 150 participants (81 females, 65 males, and 4 others) from Prolific.  We recruited normal and color-vision anomalous individuals to obtain representative results. Half the participants were asked to make judgments about the \emph{mean} value of the field. Specifically, they were prompted to ``select the image that exhibits \emph{higher} terrain on average.'' The other half was asked to judge the \emph{variance} with a prompt instructing them to select the image with the ``greater \emph{variation} in terrain.'' Participants were paid \$6. 

\vspace{1.5mm}\noindent\textbf{Experiment Design: }The experiment is a factorization of three factors: Statistic (\emph{mean} and \emph{variance}) $\times$ Colormap (4 designs) $\times$ Magnitude (2 levels: low and high). Statistic was a between-subjects factor. Colormap and Magnitude were varied within subjects. Magnitude corresponds to the baseline level of the statistic: stimuli falling in the lower half of the mean or variance range are categorized as `low',  whereas those in the upper half are `high'. We selected four colormaps to test our hypotheses (\S\ref{sec:hypotheses}): \emph{greyscale} (linear interpolation of L* $\in [0,100]$), \emph{viridis}, \emph{blue-red}, and \emph{jet} (Figure~\ref{fig:psyc_functions}).

\vspace{1.5mm}\noindent\textbf{Procedures: } Participants first saw a brief tutorial followed by 10 practice trials. They then completed four analyzed blocks corresponding to the four colormaps presented in random order. Each block consisted of 50 trials, half of which were high-magnitude (i.e., fields in the top 50th percentile of mean or variance) and the other half were low-magnitude fields. A trial consisted of a fixation cross (1 second), followed by a pair of scalar fields, displayed for 1.5 seconds (see Figure~\ref{fig:stimulus}). One of the two fields, the target, exhibited a slightly higher summary statistic than the other (the reference). The difference between the target and the reference was controlled via a staircase procedure: If the participant makes a correct guess, the difficulty of the subsequent trial is increased by reducing the difference between the two fields (in mean or variance). Conversely, we increase the difference after an incorrect response, making the next trial easier. We used a 3-down, 1-up step as is common in psychometric experiments. The step length was set to 5\% of the full statistic range, with an initial difficulty set to 30\%. The staircase procedure converges to the just-noticeable difference (JND) at which participants can reliably judge the statistic.

In addition to the analyzed trials, we inserted two engagement checks within each block. The checks consisted of stimuli with a very large difference between the target and the reference, making for an easy judgment. Participants who failed more than three checks were removed from the analysis (only 1 participant was ultimately excluded). After completing a total of 200 analyzed trials, participants were prompted to provide an open-ended description of the strategy they had followed, including any ``visual features'' they relied upon to make their judgments. Lastly, participants completed a color-vision check consisting of 15 Ishihara slides.

\subsection{Psychometric Modeling}

 We obtained a total of 29,320 binary responses (correct for identifying the true target). We used the \textbf{\texttt{quickpsy}} R package~\cite{linares2016quickpsy}, fitting the responses from each participant to four psychometric functions, one for each of the four colormaps. The fitted functions predict the probability of correctly identifying the target as a function of signal strength (i.e., the difference in mean or variance between the stimulus pairs). We used a cumulative normal distribution as the psychometric function family, setting the guess rate at $p=0.5$ and allowing for a lapse rate of up to .05.  Figure~\ref{fig:psyc_functions} illustrates the fitted functions. We then extracted the \emph{75\% JND}, which reflects the signal threshold needed to achieve a correct guess 75\% of the time (halfway between chance and perfect reliability). A lower JND indicates a better ability to estimate the ensemble statistics.

\begin{figure*}
    \centering
    \includegraphics[width=.9\linewidth]{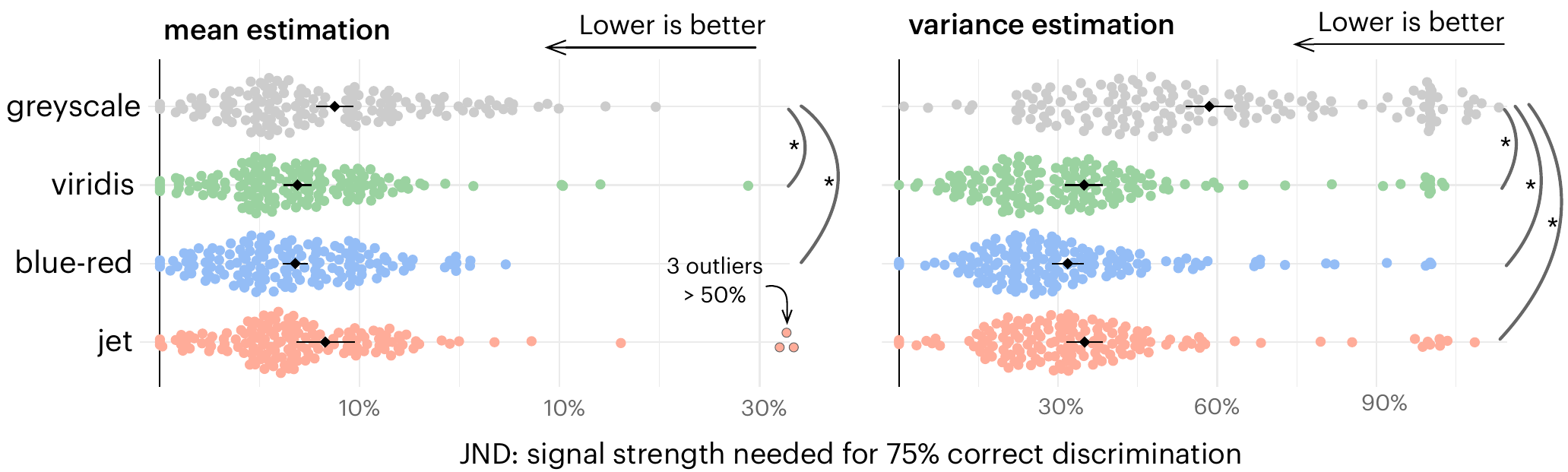}
    \caption{Just noticeable difference (JND) thresholds for the two ensemble statistics. Diamonds depict average thresholds for all participants ($\pm$95\% CIs). Dots represent JNDs for individual participants. Arcs denote significant differences between colormaps ($p<.05$, Tukey-adjusted).}
    \label{fig:jnd}
    \vspace{-4mm}
\end{figure*}

\subsection{Results}

Participants completed the experiment in 16.7 minutes on average ($\sigma=14.5$). Overall, estimating the mean required a signal difference of just 7.68\% between the two stimulus fields on average (95\% confidence intervals: 7.18--8.17\%). This was significantly lower than the amount of signal needed to discriminate variance (40.1\%, CI: 38--42.1\%). We fitted the results to a mixed-effects linear model to analyze differences in the JND thresholds as a function of colormap. The model includes fixed effects of colormap and baseline signal magnitude. We also incorporated a random intercept to model individual differences among participants. Figure~\ref{fig:jnd} illustrates JNDs for the two ensemble statistics across colormaps. 

\vspace{2mm}\noindent\textbf{Mean Estimation: }For the mean statistic, the model predicts a significant effect for the choice of colormap. Specifically, \emph{viridis} showed a significantly lower JND compared to greyscale (1.86\% advantage, $t(511)=2.824$, $p<.05$). Similarly, \emph{blue-red} also required a lower signal for successful discrimination (advantage over greyscale: 1.97\%, $t(511)=2.997$, $p<.05$). Rainbow (\emph{jet}) demonstrated similar performance to greyscale, with the average seemingly influenced by three outlying participants who exhibited unusually high JND. The baseline signal magnitude did not affect mean estimation ($t(511)=-0.387$) or interact with the colormap. 

\vspace{2mm}\noindent\textbf{Variance Estimation: } The choice of colormap appears to have a stronger impact on variance estimation. The perceptual greyscale performed worse than all three colormaps tested. Specifically, compared to greyscale, \emph{viridis} reduced the JND threshold by an average of 2.37\% ($t(518)=10.775$, $p<.0001$). Similarly, \emph{jet} also exhibited a comparable advantage over greyscale (2.35\%, $t(518)=10.693$, $p<.0001$). The advantage for \emph{blue-red} was slightly higher (2.68\%, $t(518)=12.181$, $p<.0001$), but still consistent with the other colormaps. All other comparisons were non-significant. 

\begin{figure}[t]
    \centering
    \includegraphics[width=.75\linewidth]{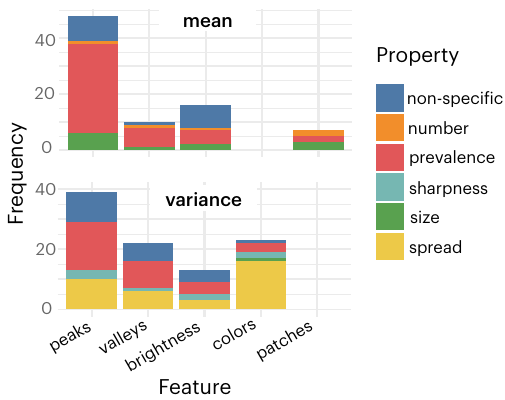}
    \caption{Visual features and properties reported by participants.}
    \label{fig:strategy}
    \vspace{-4mm}
\end{figure}

\vspace{2mm}\noindent\textbf{Perceptual Strategies: } We coded participants' self-reported strategies to understand visual features and properties used in making ensemble judgments (see Figure~\ref{fig:strategy}). When estimating the mean, participants predominantly focused on the peak color of the scale (59.2\%), often gauging the `prevalence' of that color in the displays. Many participants also noted the `number' and `size' of peaks as indicative of the overall mean. The second most common feature considered was the `brightness' of the display, although it was only referenced about 19.8\% of the time. Valleys, representing the lowest color on the scale, were the third most common feature (12.3\%). In contrast, when estimating variance, both peaks and valleys emerged as the two most commonly relied-on features (40.2\% and 22.7\%, respectively). Participants frequently assessed the `prevalence' of these extreme colors, although some also noted the `spread' as a cue for variance. Aside from the extreme ends of the scale, the spread for all `colors' was commonly reported (23.7\%). A few participants cited display `sharpness' as a proxy for variance.

\section{Discussion and Conclusion}

We evaluated people's ability to make rapid, gist-level assessments of data in pseudocolored scalar fields. When estimating the mean, a basic greyscale ramp showed lower sensitivity (higher JND) compared to multi-hue or diverging colormaps. Notably, a rainbow scale performed similarly to greyscale, which may be attributed to difficulty by participants who have color-vision deficiency.

Crucially, these judgments do not always rely on the luminance (or `brightness') of displays. Instead, extracting summaries primarily involved assessing properties of color segments resulting from the use of multi-hue, diverging, or rainbow colormaps. Although technically `artifacts'~\cite{rogowitz2001blair,borland2007rainbow}, these features appear to drive better interpretation. Indeed, participants frequently reported estimating the `prevalence', `number', and `size' of colors at the extreme ends of the scale (especially hotspots) -- summary properties that the visual system should be able to assess at a glance~\cite{franconeri2009number,ariely2001seeing,alvarez2011representing}. Our results thus lend support to H2. Although luminance was mentioned by some participants, it seemed less important for interpreting global properties. Furthermore, the fact that a luminance ramp (greyscale) was the least effective provides evidence against H1.

The variance statistic yielded comparable results with improved performance for rainbows. Participants similarly reported evaluating the prevalence of peaks and valleys, along with the overall `spread' of colors along the scale. These findings reinforce the idea that discrete color features in continuous representations may be beneficial, contrary to common assumptions~\cite{borland2007rainbow}. An important corollary is that these very features could bias estimation of visual statistics. For example, a scalar field that has unusually prominent peaks but otherwise low scalar values could lead viewers to overestimate the mean. Similar perceptual biases have been reported with other visualizations~\cite{hong2021weighted}.

In contrast to Warden et al.~\cite{warden2022visualizing}, we show that incorporating hue variation offers better support for assessing ensemble properties in fields. However, results suggest that only a minimal level of hue variation is necessary -- enough to distinguish the peaks and valleys in the data. Except for greyscale, all colormaps tested fulfill this criterion. Our results also diverge from work on the perception of local gradients, which found rainbows superior~\cite{reda2019evaluating}. In contrast, we found rainbows to be no more effective than less colorful multi-hue and diverging colormaps. One important distinction with earlier works~\cite{warden2022visualizing,reda2019evaluating} is the constrained exposure time in our study, which limited responses to judgments based on ensemble representations (as intended). A limitation of this study, however, is that we used data generated by Perlin noise. Data characteristics could influence performance, so future work should attempt to generalize our results to data with different underlying structures. 

{In sum}, we showed that viewers can quickly assess the mean and, to a lesser extent, the variance of fields. This ability has practical use in scientific visualization, enabling rapid assessment of multiple datasets, such as alternative simulation runs. To our knowledge, this study is the first to measure this rapid perceptual capacity in pseudocolor fields. Future work could explore additional summary properties like skewness, or test other displays (e.g., choropleths).

\acknowledgments{
This research was supported by the Office of Science, U.S. Department of Energy, under contract DE-AC02-06CH11357. KR was supported by the Argonne faculty sabbatical program and by NSF award 1942429.}

\bibliographystyle{abbrv-doi}

\interlinepenalty=10000
\balance

\bibliography{00_color}
\end{document}